\documentclass[aps,prb,twocolumn,groupedaddress]{revtex4-1}

\usepackage{graphicx}
\usepackage{dcolumn}
\usepackage{bm}
\usepackage{amsmath}
\usepackage{amssymb}
\usepackage{latexsym}
\usepackage{epsfig}
\usepackage{amsbsy}
\usepackage{array}
\usepackage{amssymb}
\usepackage{setspace}
\usepackage{bm}
\usepackage{subfigure}
\usepackage{caption}

\def\sint{\ifmmode{- \!\!\!\!\!\! \int}
    \else{\hbox{$- \!\!\!\! \int \ $}}\fi}

\renewcommand{\raggedright}{\leftskip=0pt \rightskip=0pt plus 0cm}

\begin{document}


\title{Charge transport properties of the Majorana zero mode induced noncollinear spin selective Andreev reflection }

\author{Xin Shang}
\author{Haiwen Liu}
\email[Haiwen Liu: ]{haiwen.liu@bnu.edu.cn}
\author{Ke Xia}
\email[Ke Xia: ]{kexia@bnu.edu.cn}
\affiliation{Department of Physics, Beijing Normal University, Beijing 100875, China}


\begin{abstract}
We study the charge transport of the spin-selective Andreev reflection(SSAR) effect between a spin polarized scanning tunneling microscope(STM) tip and  a Majorana zero mode(MZM).
Considering both the MZM and the excited states, we calculate the conductance and the shot noise power of the noncollinear SSAR using scattering theory.
We find the influence of first excited states cannot be avoided when the spin polarization direction of the STM tip and the MZM are not collinear.
In this case, the first excited states give rise to inside peaks and change the conductance peak value at zero energy.
Moreover, we numerically calculate the shot noise power and the Fano factor of the SSAR effect.
Our calculation shows that the shot noise power and the Fano factor are related to the angle between the spin polarization direction of the STM tip and that of the MZM.
These transport properties of the SSAR effect provide additional characteristics to detect the MZM via SSAR.
\end{abstract}



\maketitle

\section{Introduction}  
The Majorana  zero mode (MZM) is a special type of Bogoliubov quasiparticle excitation
with non-Abelian statistics, which forms the base of topological quantum
computations\cite{MOORE1991362,PhysRevB.61.10267,1063-7869-44-10S-S29,KITAEV20032,RevModPhys.80.1083,PhysRevLett.105.177002}, and has received  a large
amount of research interest since being proposed.
There are a number of methods that have been used to generate and detect the MZM in condensed matter systems\cite{RevModPhys.80.1083,RevModPhys.87.137,Aguado2017Majorana};these include a chiral p-wave superconductor\cite{PhysRevB.73.220502}, the $\nu=5/2$ fractional quantum Hall
system\cite{MOORE1991362,PhysRevB.61.10267}, topological insulator(TI)/s-wave
superconductor(SC) interfaces with the MZM in the vortex
core\cite{PhysRevLett.100.096407}, and proximity-induced superconductors for spin-orbit coupled nanowires\cite{PhysRevLett.105.177002,PhysRevLett.105.077001}.
An electron with its spin direction aligned with the MZM  will undergo an Andreev reflection, while an electron with opposite spin direction will not\cite{PhysRevLett.112.037001,PhysRevLett.112.217001,PhysRevLett.115.177001,PhysRevLett.116.257003,PhysRevB.94.224501};
this allows us to detect the existence of the MZM using ferromagnetic STM\cite{PhysRevLett.112.037001,PhysRevLett.112.217001,PhysRevLett.115.177001,PhysRevLett.116.257003,PhysRevB.94.224501}.

However, previous studies of 2D SSAR have only the collinear spin transport, for example, the spin polarization of the lead being parallel or antiparallel to that of the MZM.
The excited states, which have low energy above the MZM, were completely ingnored\cite{PhysRevB.94.224501}.
To obtain more information concerning SSAR, we consider both the MZM and the excited state in the SSAR effect when the spin polarization direction of the STM tip and the MZM are noncollinear.
In addition, compared to the conductance, the shot noise may 
shed more light on the underlying physical properties of the system\cite{BLANTER20001}.
In particular, the shot noise can be used to determine the charge and statistics of the quasiparticles relevant to transport, and to reveal information concerning the potential and internal energy scales of mesoscopic systems\cite{BLANTER20001}.
In this paper, we calculate the shot noise power and the Fano factor of the SSAR to provide additional approaches to detect the MZM using SSAR.

The paper is organized as follows.
In Section~\ref{model}, we build a model to calculate the conductance and the shot noise power using Green's function and the Fisher--Lee relation.
In Section~\ref{conductance}, we discuss the conductance of the noncollinear SSAR effect with different $\theta$,i.e.with different angles between the spin polarization of the FM tip
and that of the MZM, and that of different coupling between the STM and the vortex core at different temperatures.
In Section~\ref{shotnoise}, we discuss the shot noise power and the Fano factor of noncollinear SSAR effect.
\\
\begin{figure}
 \scalebox{0.4}[0.4]{\includegraphics[12,10][500,425]{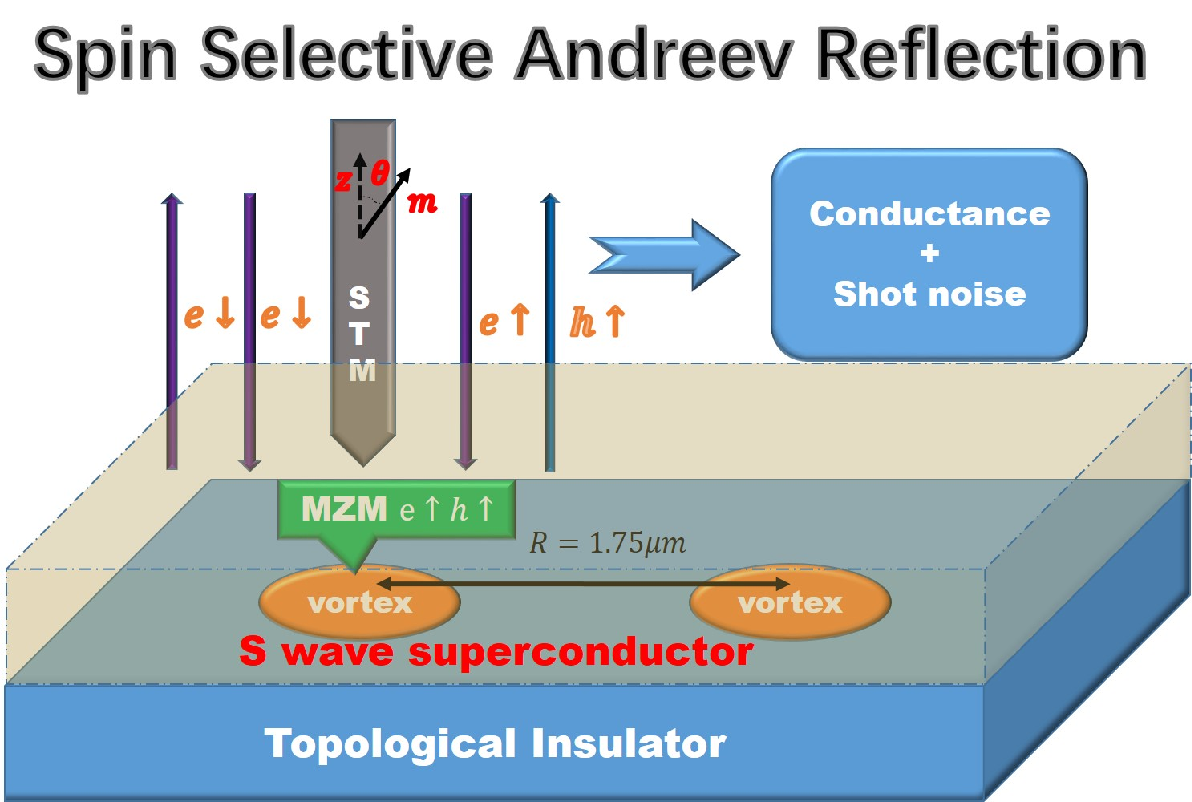}}
 \caption{\raggedright The Majorana induced Spin Selective Andreev Reflection.
 A MZM exists in the vortex core of a 3D topological insulator, which is under an s-wave superconductor.
 An electron with the same spin as the MZM  will undergo an Andreev reflection, whereas an electron with an opposite spin direction will not. The conductance and the shot noise power of this SSAR induced current may relate to the spin polarization of the STM tip.
  \label{fig1}}
\end{figure}

\section{Model}\label{model}
As shown in Figure ~\ref{fig1}, we consider a topological insulator(TI) covered by a superconductor.
Superconductivity is induced in the TI via the proximity effect , and a vortex state is formed in the surface of the TI under a magnetic field.
At the center of the vortex core(r = 0), the spin polarization of the MZM is parallel to the magnetic field.

Let us construct the Hamiltonian of the vortex state in a topological superconductor(TS).
This TS is modeled by a helical surface state with Rashba spin--orbit coupling and  proximity-induced superconductivity\cite{PhysRevB.94.224501}.
The helical surface state is the surface state of a 3D topological insulator in the x--y plane.
We can generalize the Hamiltonian in the x--y plane to a spherical surface of radius R.
The single-electron Hamiltonian of a helical surface state is\cite{PhysRevB.94.224501}:
\begin{equation}
H_{0e}=-\frac{\alpha}{R\hbar}\hat{L}\times\hat{\sigma}-\mu,
\end{equation}
where $\alpha$ is the spin-orbit coupling strength, $\hat{\sigma}$ is the Pauli matrices, $\hat{L}$ are the orbital angular momentum, and $\mu$ is the chemical potential.

The Hamiltonian of the proximity-induced superconducting state can be described as:
\begin{equation}
H_{\Delta}=\Delta(c_{\downarrow}c_{\uparrow}-c_{\uparrow}c_{\downarrow})+\Delta^*(c_{\downarrow}^+c_{\uparrow}^+-c_{\uparrow}^+c_{\downarrow}^+),
\end{equation}
where $\Delta$ is the proximity induced superconducting order parameter ans $c^{(+)}_{\sigma}$ is the electron annihilation (creation) operator with $\sigma=\uparrow\downarrow$ denoting the spin.

Then, in the standard Nambu representation, the field operator can be defined as
\begin{equation}
\hat{\psi}(r)\equiv[\hat{c}_{\uparrow}(r),\hat{c}_{\downarrow}(r),\hat{c}^+_{\downarrow}(r),-\hat{c}^+_{\uparrow}(r)]^T.
\end{equation}

The Hamiltonian of the proximity-induced superconducting state in then
\begin{equation}
H_{\Delta}=
\left[
 \begin{matrix}
   0 & \Delta I\\
   \Delta^*I & 0\\
  \end{matrix}
  \right],
\end{equation}
where $I$ is the  unit matrix.

The Hamiltonian of the vortex state can be written as:
\begin{equation}
H_{v}=
\left[
 \begin{matrix}
   H_{0e} & \Delta I\\
   \Delta^*I & H_{0h}\\
  \end{matrix}
  \right],
\end{equation}

where, $H_{0h}$ is the single-hole Hamiltonian defined as $-\sigma_yH_{0e}\sigma_y$ and $\sigma_y$ is a pauli matrix.

The vortex state can be described as $\Delta=\Delta(\theta)e^{i\phi}$. Here the factor $e^{i\phi}$ describes a vortex with a winding number of 1 and $\Delta(\theta)=\Delta_0\tanh{(\frac{R\sin\theta}{\xi_0})}$, where $\xi_0$ characterizes the size of the vortex core.

By diagonalizing  $H_{v}$, we can define a new quantum number of $K_z$\cite{PhysRevB.94.224501} where $K_z|\Phi_m>=m|\Phi_m>$, where m and $|\Phi_m>$ are the eigenvalue and eigenfunction of $K_z$. Here $K_z\equiv l_z+\sigma_z+\tau_z$ with $l_z$ is the orbital quantum number in the z direction, $\sigma$ is the spin quantum number in the z direction, $\tau_z$ is the spin-orbit-pseudospin quantum number referring to the particle-hole degree of freedom and
$|\Phi_m>$ is the four-component wave function\cite{PhysRevB.94.224501}.
\begin{equation}
|\Phi_m>=[e^{im\phi}u_1,e^{i(m+1)\phi}u_2,e^{i(m-1)\phi}v_1,e^{im\phi}v_2]^T
\end{equation}
where m is the eigenvalue of $K_z$.

The eigenvalue problem then becomes
\begin{equation}
H_{tot} |\Phi_m>=E_m|\Phi_m>.
\end{equation}
The four-component eigenfunction basis\cite{PhysRevB.94.224501,PhysRevB.82.214509} in $|\Phi_m>$ can be expressed in terms of the spherical harmonic functions:
$e^{im\phi}u_1(m)=\sum_la_lY_l^m$,
$e^{i(m+1)\phi}u_2(m+1)=\sum_lb_lY_l^{m+1}$,
$e^{i(m-1)\phi}v_1(m-1)=\sum_lc_lY_l^{m-1}$,
$e^{im\phi}v_2(m)=\sum_ld_lY_l^m$,
 where
 $Y_l^m(\theta,\phi)=P_l^m(\cos\theta)e^{im\phi}/\sqrt{2\phi}$ and $P_l^m(\cos\theta)$ is the associated Legendre polynomial.

By directly diagonalizing the $H_{v}$, we can obtain the wave function and the energy spectrum.
In our numerical calculations, we set $R=50\xi_0$, $\alpha=30$ meV$\cdot$nm, $\xi_0$=35 nm, $\Delta_0=1$ meV, and $\mu=90$ meV, which are comparable to the experiment data in $Bi_2Se_3$\cite{PhysRevLett.114.017001}.
Taking a cutoff in the orbital angular momentum $l$ to be approximately 200
we find that for m = 0, $E_0$ = zero (numerically $\pm4\times10^{-6}$ meV).
Here $u1=v2\ne0$, and $u2=v1=0$; this means that a spin up electron and a spin up hole occupy the MZM.
For the $m=1$ state, $E_{-1}=-0.06$ meV, $v1\ne0, u2=0, u1=0$, and $v2=0$; only a hole with a down spin can occupy this state.
Meanwhile, for $m=-1$, $E_1=0.06$ meV, $u2\ne0$, and u1, v1, and v2=0, only a spin down electron can occupy this state.
These are the first excited states of the vortex.
When $|m|>1$,u1, u2, v1, and v2 are equal to zero at the core of the vortex.

Next, let us consider the total Hamiltonian of a system with a vortex state coupling to a spin polarization STM tip.
The Hamiltonian of the electron on the STM tip can be described as
\begin{equation}
\begin{split}
H_{L,e}=\sum_{\sigma}\hat{d}_{L,\sigma}^{+}(\varepsilon_{\sigma}-\mu_{L})\hat{d}_{L,\sigma}\\
+\sum_{\sigma\sigma'}\hat{d}_{L,\sigma}^{+}[\vec{M}\cdot\vec{\sigma }]_{\sigma\sigma'}\hat{d}_{L,\sigma'},
\end{split}
\end{equation}
where $\hat{d}_{L,\sigma}^{(+)}$ denotes the electron annihilation (creation) operator of theSTM tip with $\sigma$ spin, $\mu_{L}$ indicates the chemical potential of the STM tip (set to zero), $\varepsilon_{\sigma}$ is the kinetic energy of the STM tip with $\sigma$ spin, $\vec{M}$ is the spin related potential and $\vec{\sigma }$ are Pauli matrices.

The Hamiltonian of the STM tip is
\begin{equation}
H_{L}=
\left[
 \begin{matrix}
   H_{L,e} & 0\\
   0 & H_{L,h}\\
  \end{matrix}
  \right],
\end{equation}
where $H_{L,h}=-\sigma_{y}H_{L,e}\sigma_{y}^*$ is the Hamiltonian of the hole on the STM tip.

The coupling between the vortex states and the STM tip can be described using the following Hamiltonian(the STM tip is in contact with the vortex at the 0 site):
\begin{equation}
\begin{split}
H_{v-s}=\sum_{\sigma}\{2t_c\hat{c}_{\sigma}^+\hat{d}_{L,\sigma}+H.c.\},
\end{split}
\end{equation}
where $t_c$ is the coupling strength between the vortex and the STM tip.

The total Hamiltonian of the system is
\begin{equation}
H_{tot}=
\left[
 \begin{matrix}
   H_{L} & t_c I\\
   t_c I & H_{v}\\
  \end{matrix}
  \right].
\end{equation}

The retarded Green¡¯s function of the system can be obtained via Dyson's equation:
\begin{equation}
G^{tot}=\frac{1}{(G^{(0,R)})^{-1}-\Sigma^{r}}.
\end{equation}
Here, the single-particle retarded Green¡¯s function $G^{0R}$ can be constructed with the wave functions $\hat{\Psi}_m$ and the eigenvalue $(E_m)$ of the vortex state:
\begin{equation}
G^{(0,R)}=\sum_m\frac{|\Psi_m><\Psi_m|}{E-E_m+i\delta},
\end{equation}
where $\hat{\Psi}_m=[e^{im\phi}u_1,e^{i(m+1)\phi}u_2,e^{im\phi}v_2,e^{i(m-1)\phi}v_1]^T$ and $\delta$ is a positive infinitesimal.

We assume that the spin polarization of the FM tip $\hat{M}_{L}$ has an angle $\theta$ with respect to the z direction(the direction of the MZM spin polarization).

The self-energy $\Sigma^{r}=t_c^2\lambda^{r}$, where $\lambda^{r}=\sum_m \frac{|\phi^1_m><\phi^1_m|}{E-E_m+i\delta}$, is the single-particle retarded Green¡¯s function of the STM tip. $E_m$ is the eigenvalue of $H_{L}$ and $|\phi^1_m>=[\phi_1,\phi_2,\phi_4,\phi_3]^T$, where $[\phi_1,\phi_2,\phi_4,\phi_3]^T$ is the eigenfunction of $H_{L}$).

Specifically, when $\theta=0^0$, $\lambda^{r}$ is a diagonal matrix with elements $\lambda_{e\uparrow},\lambda_{e\downarrow},\lambda_{h\uparrow}$, and $\lambda_{h\downarrow}$.
In this case, $\lambda_{e\uparrow}$ and $\lambda_{e\downarrow}$ are the densities of states of the spin up and spin down electrons, while $\lambda_{h\uparrow}$ and $\lambda_{h\downarrow}$ are the densities of states of the spin-up and spin-down holes obtained by wide-band limit.

Then, the general form of $\lambda^{r}$ is
\begin{equation}
\lambda^{r}=
\left[
 \begin{smallmatrix}
 \setlength{\arraycolsep}{0.1pt}
   \lambda_{e\uparrow\uparrow} & \lambda_{e\uparrow\downarrow} & 0 & 0\\
   \lambda_{e\downarrow\uparrow} & \lambda_{e\downarrow\downarrow} & 0 & 0\\
   0 & 0 & \lambda_{h\uparrow\uparrow} & \lambda_{h\uparrow\downarrow} \\
   0 & 0 & \lambda_{h\downarrow\uparrow} & \lambda_{h\downarrow\downarrow}
  \end{smallmatrix}
  \right]
  \label{self-energy},
\end{equation}
where $\lambda_{e\uparrow\uparrow}=\lambda_{e\uparrow}\frac{1+\cos(\theta)}{2}+\lambda_{e\downarrow}\frac{1-\cos(\theta)}{2}$,
   $\lambda_{e\uparrow\downarrow}=(\lambda_{e\uparrow}-\lambda_{e\downarrow})\frac{\sin(\theta)}{2}$,
   $\lambda_{e\downarrow\uparrow}=(\lambda_{e\uparrow}-\lambda_{e\downarrow})\frac{\sin(\theta)}{2}$,
   $\lambda_{e\downarrow\downarrow}= \lambda_{e\downarrow}\frac{1+\cos(\theta)}{2}+\lambda_{e\uparrow}\frac{1-\cos(\theta)}{2}$,
   $\lambda_{h\uparrow\uparrow}= \lambda_{h\uparrow}\frac{1+\cos(\theta)}{2}+\lambda_{h\downarrow}\frac{1-\cos(\theta)}{2}$,
   $\lambda_{h\uparrow\downarrow}= (\lambda_{h\uparrow}-\lambda_{h\downarrow})\frac{\sin(\theta)}{2}$,
   $\lambda_{h\downarrow\uparrow}= (\lambda_{h\uparrow}-\lambda_{h\downarrow})\frac{\sin(\theta)}{2}$, and
   $\lambda_{h\downarrow\downarrow}=\lambda_{h\downarrow}\frac{1+\cos(\theta)}{2}+\lambda_{h\uparrow}\frac{1-\cos(\theta)}{2} $.

In our calculation, we use the parameters as $\lambda_{e\uparrow}=-0.125-0.08i$,$\lambda_{e\downarrow}=0.05 \lambda_{e\uparrow}$, $\lambda_{h\uparrow}=0.125-0.08i$, and $\lambda_{h\downarrow}=0.05\lambda_{h\uparrow}$.
In this case, the STM is nearly fully polarized. The S matrix can be obtained via the Fisher-Lee formula\cite{PhysRevB.23.6851}:
\begin{equation}
S=-I+i\Gamma^{1/2}\times G^{tot}\times \Gamma^{1/2},
\end{equation}
where, $\Gamma$ is the broadening function, which is defined as $\Gamma=i(\Sigma^{r}-\Sigma^{r+})$.
The S matrix is a $4 \times 4$ matrix :
\begin{equation}
S=
\left[
 \begin{matrix}
 \setlength{\arraycolsep}{0.2pt}
   r_{ee} & r_{eh} \\
   r_{he} & r_{hh} \\
  \end{matrix}
  \right],
\end{equation}
where $r_{ee}(r_{hh})$ is a $2 \times 2$ matrix describing the probability of a electron(hole) being reflected as a electron(hole), while $r_{eh}(r_{he})$ is a $2 \times 2$ matrix describing the probability of a electron(hole) being reflected as a hole(electron) in spin space.
The current $I_c$ can be defined as
\begin{equation}
\begin{split}
I_c=\frac{e}{h}\int_{0}^\infty[<a^+_e(E)a_e(E)>-<b^+_e(E)b_e(E)>\\
-<a^+_h(E)a_h(E)>+<b^+_h(E)b_h(E)>]dE,
\end{split}
\end{equation}
where $a^{+}_{e(h)}$ is the generate (annihilation) operator of an incoming electron(hole),
 $b^{(+)}_{e(h)}$ is the generate (annihilation) operator of a outgoing electron(hole).
The differential conductance can also be obtained\cite{Datta}.

The shot noise includes additional information concerning the fluctuation and can be calculated as\cite{BLANTER20001}
\begin{equation}
Sp(t-t')=\frac{1}{2}<\Delta I_L(t)\Delta I_L(t')+\Delta I_L(t')\Delta I_L(t)>,
\end{equation}
where $\Delta I_L(t)=I_L(t)-I_{L0}$ and $I_{L0}$ is the average of $I_L$.

The shot noise under the zero temperature limit can be obtained as follows.

When eV$>$0,
\begin{equation}
\begin{split}
Sp1=\frac{2e^3V}{h}[(r_{ee}^+a^+_e(E)r_{ee}a_e(E))(r_{eh}^+a^+_h(E)r_{eh}a_h(E))\\
-(r_{ee}^+a^+_e(E)r_{he}a_e(E))(r_{hh}^+a^+_h(E)r_{eh}a_h(E))\\
-(r_{he}^+a^+_e(E)r_{ee}a_e(E))(r_{eh}^+a^+_h(E)r_{hh}a_h(E))\\
+(r_{he}^+a^+_e(E)r_{he}a_e(E))(r_{hh}^+a^+_h(E)r_{hh}a_h(E))]\\
\end{split}.
\label{shotnoise-ev>0}
\end{equation}
When eV$<$0, $a^{(+)}_e$ should change to $a^{(+)}_h$ and  $r_{e(h)e(h)}^{(+)}$ should change to $r_{h(e)h(e)}^{(+)}$.
This is because the carrier of the charge current is the change from electron to hole.
When eV=0, the shot noise power should be $\frac{1}{2}(Sp1(eV>0)+Sp1(ev<0))$.
In other words, the shot noise power is 0 at zero temperature.

When the spin polarization of the STM tip is parallel or antiparallel to the MZM, the shot noise power\cite{BLANTER20001} can be simplified as $Sp$:
 \begin{equation}
 Sp=\frac{8e^3V}{h}r_{eh}^+r_{eh}r_{ee}^+r_{ee}.
 \end{equation}
The Fano factor\cite{BLANTER20001} is defined as
\begin{equation}
F=Sp/2eI.
\end{equation}
Both the shot noise power and the Fano factor can be obtained from the S-matrix.
\\

\section{Conductance of noncollinear SSAR}\label{conductance}
\begin{figure}
\centering
\scalebox{0.08}[0.08]{\includegraphics[276,88][2832,2484]{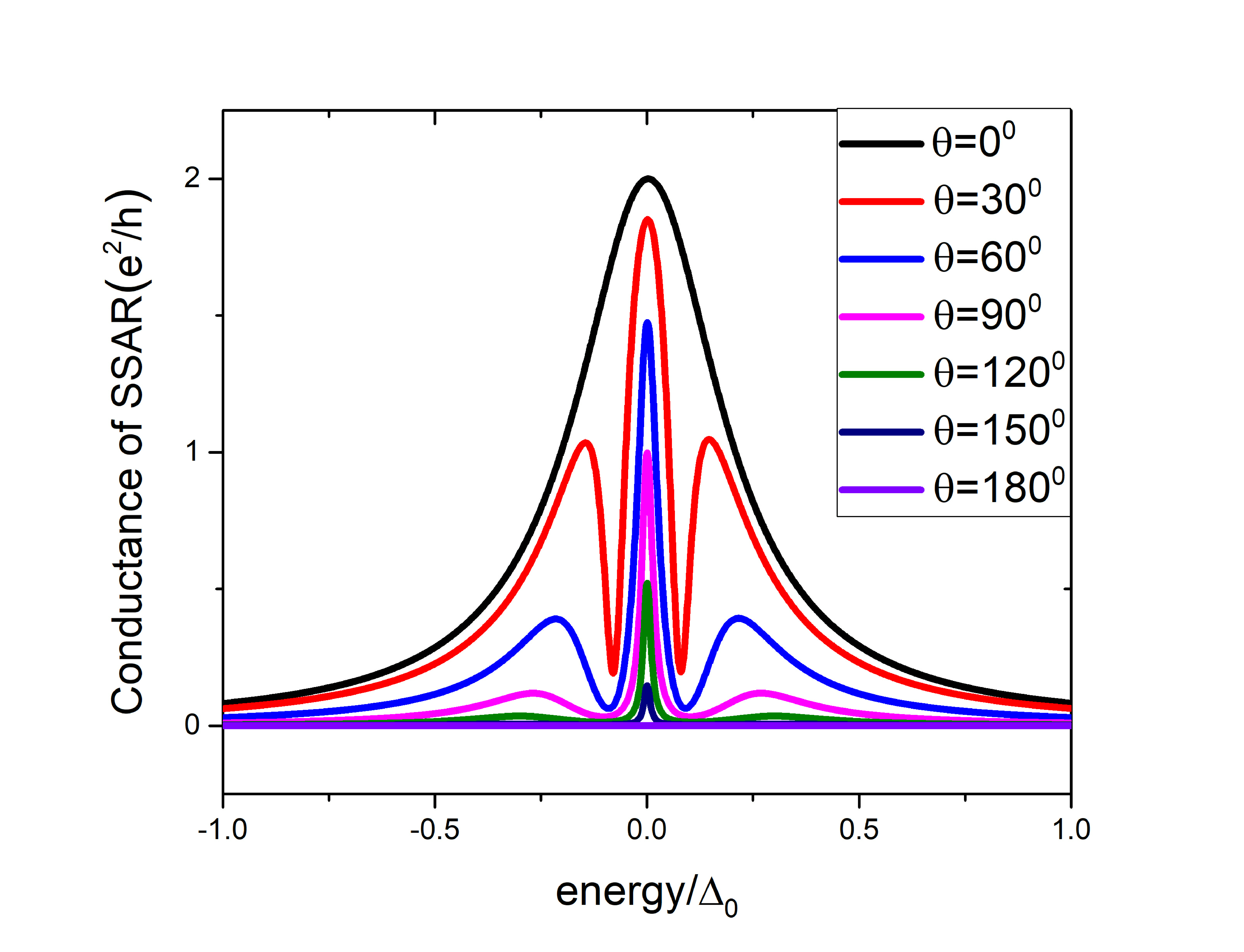}}
 \caption{\raggedright The angular dependence of the conduction in the SSAR effect. The maximum value of the conductance at zero energy is $1-\cos\theta$.
  This value is the same as two times the ratio of the electron spin projected in the direction of MZM spin polarization.
  In addition, when $\theta \ne 0^0$ or $180^0$, there are two more peaks in the conductance line.
  These two peaks maybe  due to the contribution of the first excited states.
  \label{fig2}}
\end{figure}
As shown in Figure ~\ref{fig2},
when $\theta=0^0$, the conductance line has only one broad peak with a maximum value of 2 at zero energy.
This is because only the electron at zero energy  will be reflected by the MZM as a hole.
When $\theta=180^0$, there is no conductance at all because there are no electrons that have the same spin polarization as the MZM.

However, in the noncollinear case, the maximum value of the conductance is equal to $1+\cos\theta$ (the ratio of electron that the spin direction is aligns with the MZM).
In addition, there are two more inside peaks in the conductance line.

In this case, the position of the valley between the zero-energy peak and the nearest peak on the conductance line is nearly equal to the eigenenergy of the first excited states.
To determine the influence of the first excited states, we calculated the conductance with different $\theta$ for two cases.
In the first case, we ignore the first excited states, and in the second case, we considered the first excited states.
\begin{figure}
\scalebox{0.08}[0.08]{\includegraphics[276,88][2832,2224]{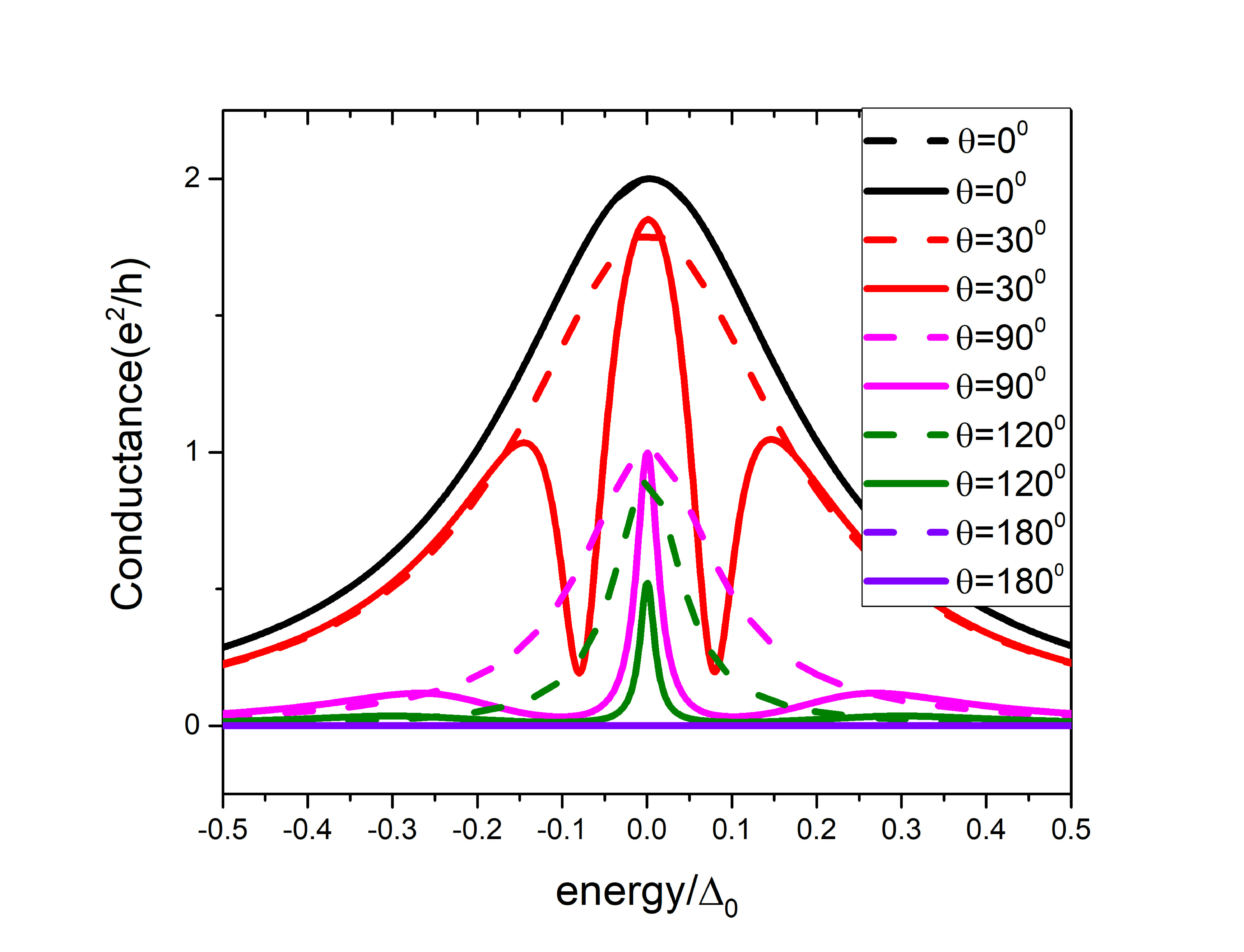}}
 \caption{\raggedright The conductance when considering the first excited states (solid line) or not (dashed line).
 The first excited states can increase the conductance when $\theta<90^0$ and reduce the conductance when $\theta>90^0$ at zero energy as well as inducing two more peaks far from the zero energy.
  \label{fig3}}
\end{figure}

From Figure ~\ref{fig3}, we can see that the first excited states have no influence on the conductance of the SSAR when $\theta=0^0$ .
However, in the noncollinear condition, the first excited states have several novel features with different $\theta$.
First, when $\theta \leqslant 90^0$, the first excited states can increase the conductance at zero energy; meanwhile, when $\theta > 90^0$, the first excited states can decrease the conductance.
Second, the first excited states can induce two additional valleys at the eigenenergy of the first excited states.

The different properties between the two conditions can be explained using the Green's function of system.
From the wave function of the vortex states, we find that only the first term(u1 for the spin-up electron, $e\uparrow$) and the third term (v2 for the spin-up hole, $h\uparrow$) of the wave function($|\Psi_0>$) have non-zero values when m = 0.
However, when m = 1, only the fourth term (for the spin-down hole, $h\downarrow$) of the wave function ($|\Psi_{-1}>$) has a non-zero value, while for m = -1, only the second term (for the spin-down electron, $e\downarrow$) of the wave function ($|\Psi_{1}>$) has a non-zero value.

Let us focus on the single-particle retarded Green¡¯s function of the vortex state $G^{(0,R)}$:
\begin{equation}
G^{(0,R)}=\sum_m\frac{|\Psi_m><\Psi_m|}{E-E_m+i\delta}=\left[
 \begin{smallmatrix}
 \setlength{\arraycolsep}{0.1pt}
   g_{e\uparrow e\uparrow}^{0,r} & 0 & g_{h\uparrow e\uparrow}^{0,r} & 0\\
   0 & g_{e\downarrow e\downarrow}^{-1,r} & 0 & 0\\
   g_{e\uparrow h\uparrow}^{0,r} & 0 & g_{h\downarrow h\downarrow}^{0,r} & 0 \\
   0 & 0 & 0 & g_{h\downarrow h \downarrow}^{1,r}
  \end{smallmatrix}
  \right].
\end{equation}
Here, $g_{e\uparrow e\uparrow}^{0,r}=\frac{u1u1}{E-E_0+i\delta}$, $g_{h\uparrow e\uparrow}^{0,r}=\frac{v2u1}{E-E_0+i\delta}$, $g_{e\uparrow h\uparrow}^{0,r}=\frac{u1v2}{E-E_0+i\delta}$, and $g_{h\downarrow h\downarrow}^{0,r}=\frac{v2v2}{E-E_0+i\delta}$ are the four components of the density of states of the MZM.
The terms $g_{e\downarrow e\downarrow}^{-1,r}=\frac{u2u2}{E-E^{-}_1+i\delta}$ and $g_{h\downarrow h \downarrow}^{1,r}=\frac{v1v1}{E-E_1+i\delta}$ are the densities of states of the m = -1 and m = 1 states respectively.

This means that the MZM is only local at the spin up channel of the hole and the electron, while the two first excited states are local at the spin-down channels of the electron and the hole, respectively.
In the collinear case, the self-energy is a diagonal matrix and there is no coupling between the two spin channels.
Only the coupling between the electron and the hole is local at the spin-up channel of the MZM.
However, in the noncollinear case, the self-energy is not diagonal and the two spin channels of the STM tip, as well as the MZM and the first excited states will be coupled together.
Therefore, the first excited states can contribute to the conductance in the noncollinear case.
\begin{figure}
\scalebox{0.08}[0.08]{\includegraphics[276,88][2832,2484]{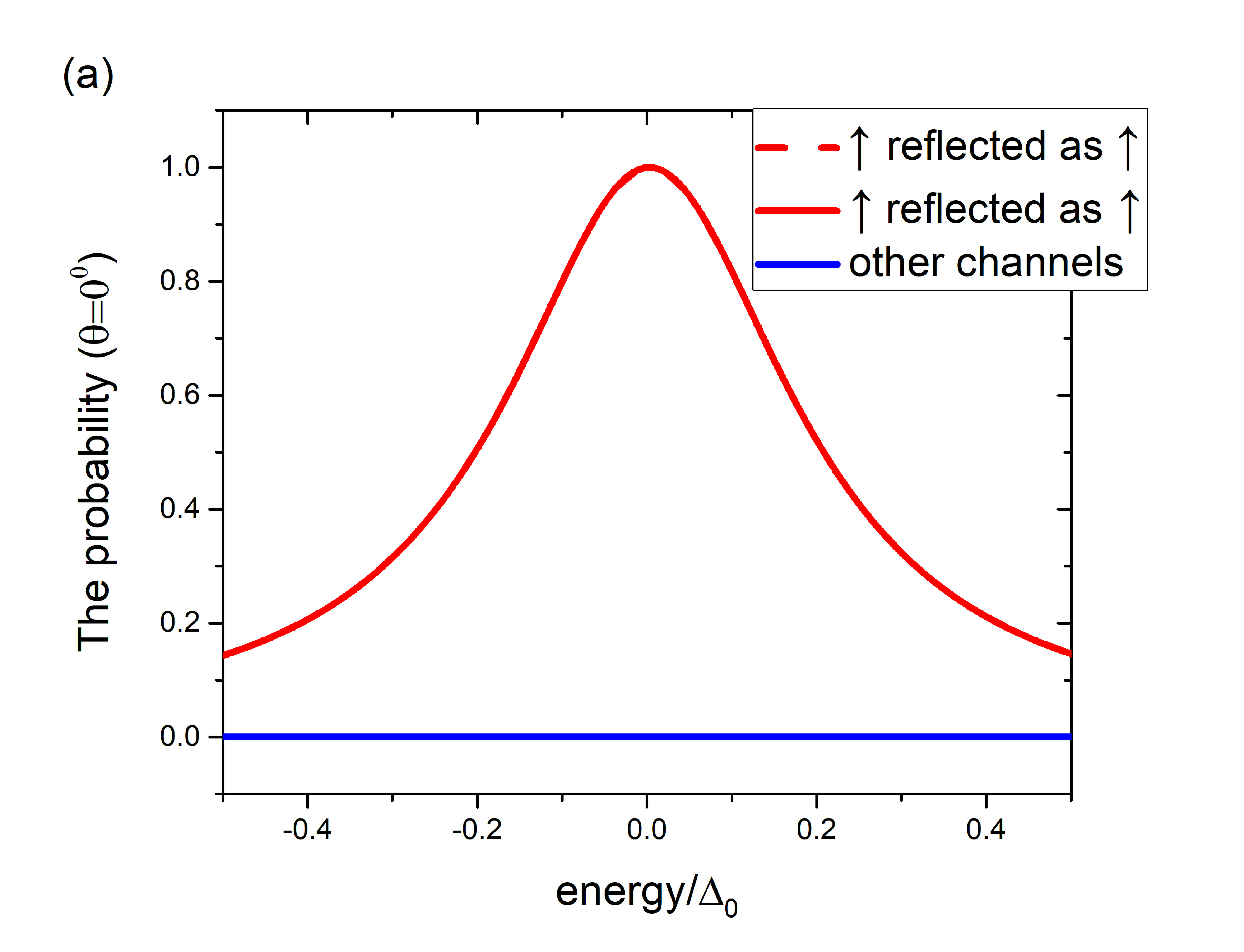}}
\scalebox{0.08}[0.08]{\includegraphics[276,88][2832,2484]{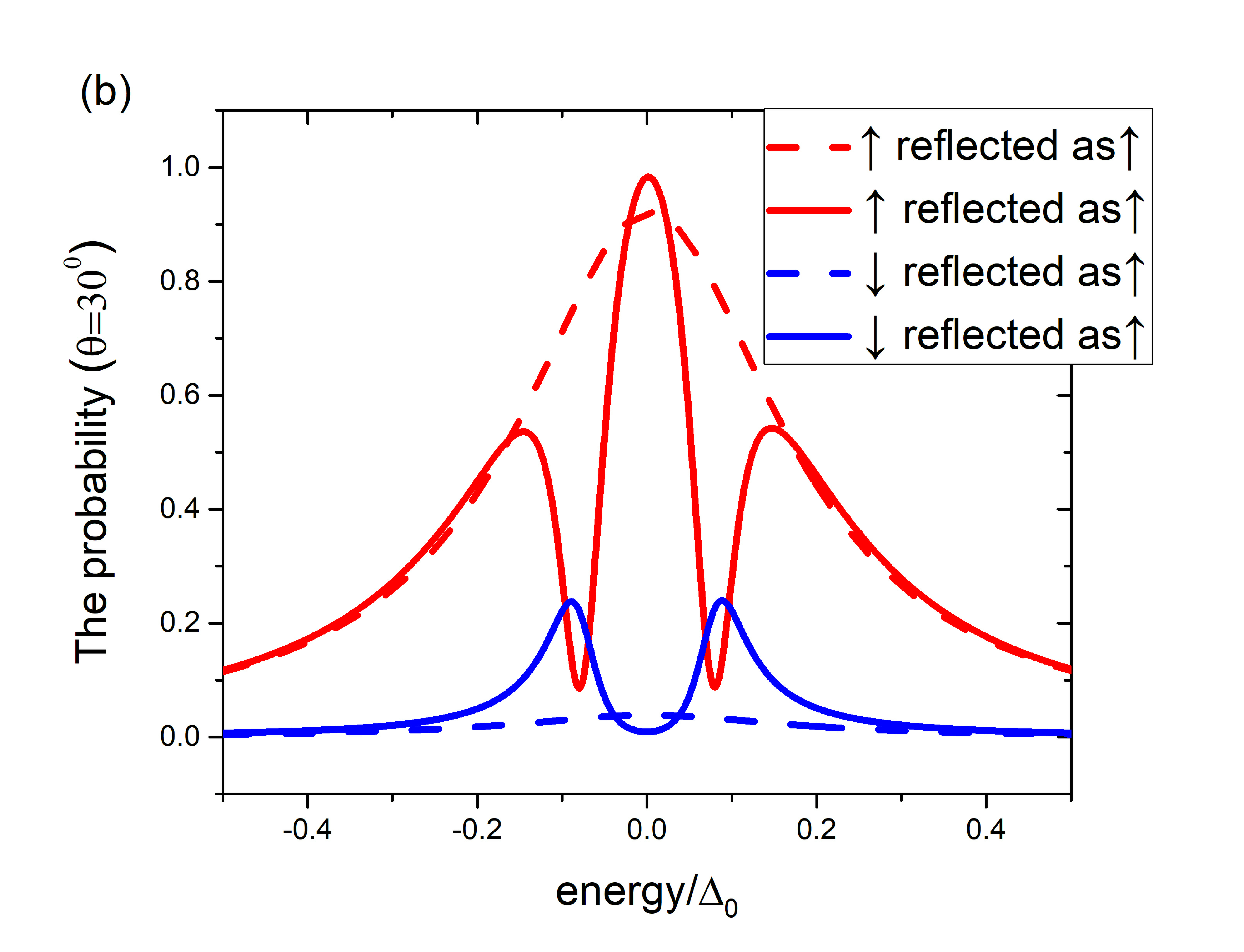}}
 \caption{\raggedright  The probability of an electron being reflected as a hole with $\uparrow\downarrow$ denoting the spin polarization when considering the first excited states (solid line) or ignoring the first excited states (dashed line). When $\theta=0^0$ (a), the first excited states cannot affect the probability of an electron being reflected as a hole. However, in the noncollinear case, such as $\theta=30^0$ (b), the first excited states can increase the probability of a spin-up electron being reflected as a hole with the same spin and reduce the probability of other channels at zero energy.
  \label{fig4}}
\end{figure}

These first excited state-induced novel features in the noncollinear case can be explained as follows.:
First, the conductance is related to the probability of an electron being reflected as a hole.
Then, as shown in Figure ~\ref{fig4}, the first excited states can increase the probability of a spin up electron being reflected as a hole and decrease the other channels at zero energy.
At zero energy, the first excited states can reduce the probability of an electron being reflected as a hole with opposite spin.
In addition,, the first excited states can reduce the probability of a spin up electron being reflected as a hole with the same spin near the energy of the first excited states.

Moreover, the first excited states can increase the probability of a spin down electron being reflected as a hole with opposite spin far from the zero energy.
Note that this influence becomes smaller when $\theta$ increase.
However, the first excited states can reduce the probability of a spin-up electron being reflected as a spin-down hole and that of a spin-down electron being reflected as a spin-down hole.
\begin{figure}
\scalebox{0.08}[0.08]{\includegraphics[276,88][2832,2484]{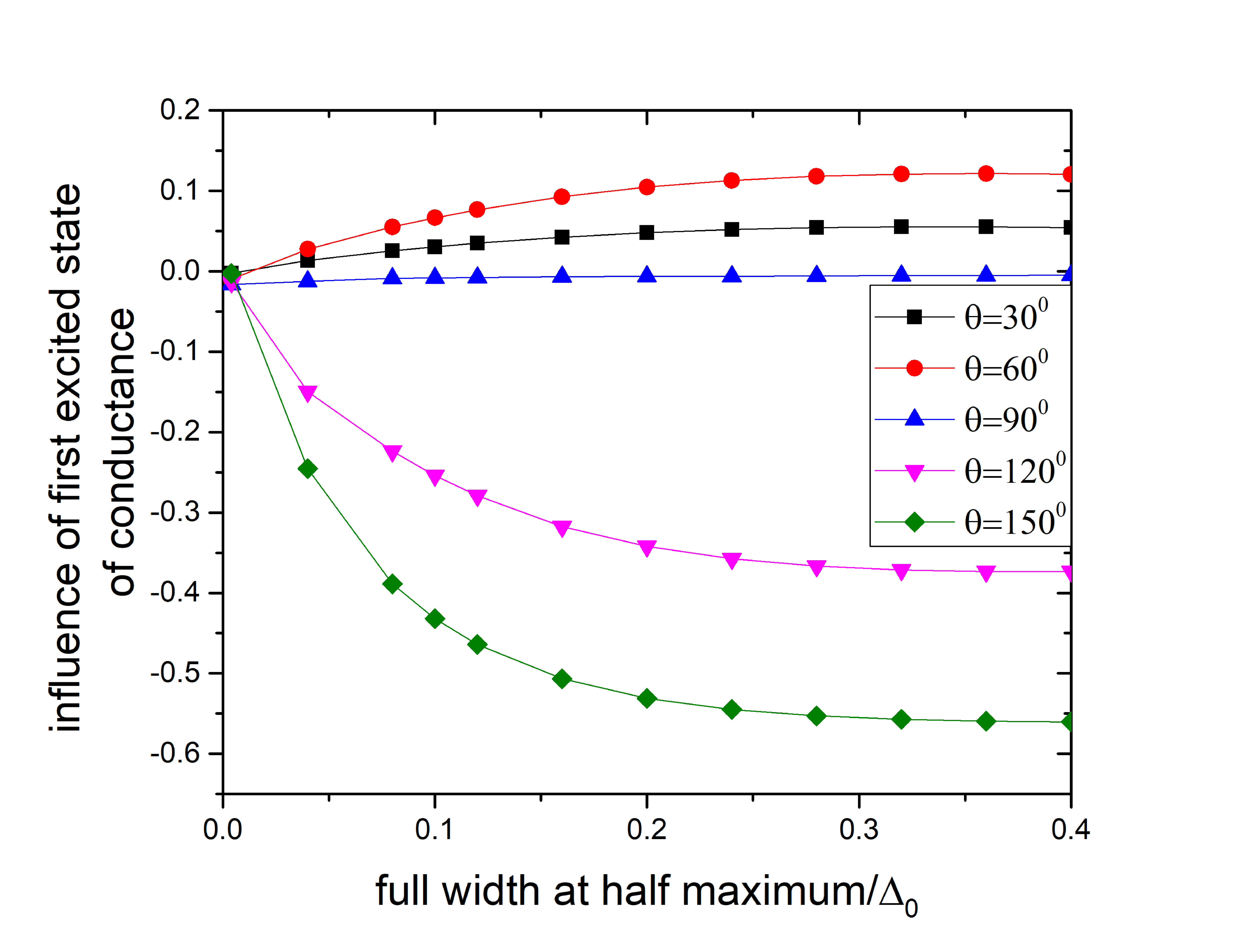}}
 \caption{\raggedright Influence of the first excited states on the conductance at zero energy when the SSAR occurs.
 As the conductance line width decrease, the influence of the first excited states decreases.
  \label{fig5}}
\end{figure}

As mentioned before, the proportion of the spin-up electrons on the FM tip is $\frac{(1+cos\theta)}{2}$.
This means that the first excited states can increase the conductance at zero energy and reduce the conductance near the energy of the first excited states when $\theta<90^0$ (when the number of spin-up electrons is greater than the number of spin-down electrons).
However, the first excited states can reduce the conductance when $\theta>90^0$ (when the number of spin-up electrons is less than the number of spin-down electrons).
When $\theta=90^0$, the first excited states have no effect on the conductance because the number of spin-up electrons is equal to spin-down electrons.

To separate the effect of the first excited states, we focus on the energy broadening of the conductance which can be influenced by $t_c$.
Here, we define the influence of the first excited state as the conductance ignoring the first excited states minus the conductance considering the first excited states.

To calculate this influence, we change the full width at half maximum as well as the energy broadening of the conductance by changing $t_c$.
As shown in Figure ~\ref{fig5}, we find that if the full width at half maximum as well as the coupling strength is very small, the effect of the first excited states is weaker.
Note that the influence of the first excited states cannot be ignored at the energy resolution of the STM(0.1 meV).
\begin{figure}
\scalebox{0.08}[0.08]{\includegraphics[276,88][2832,2484]{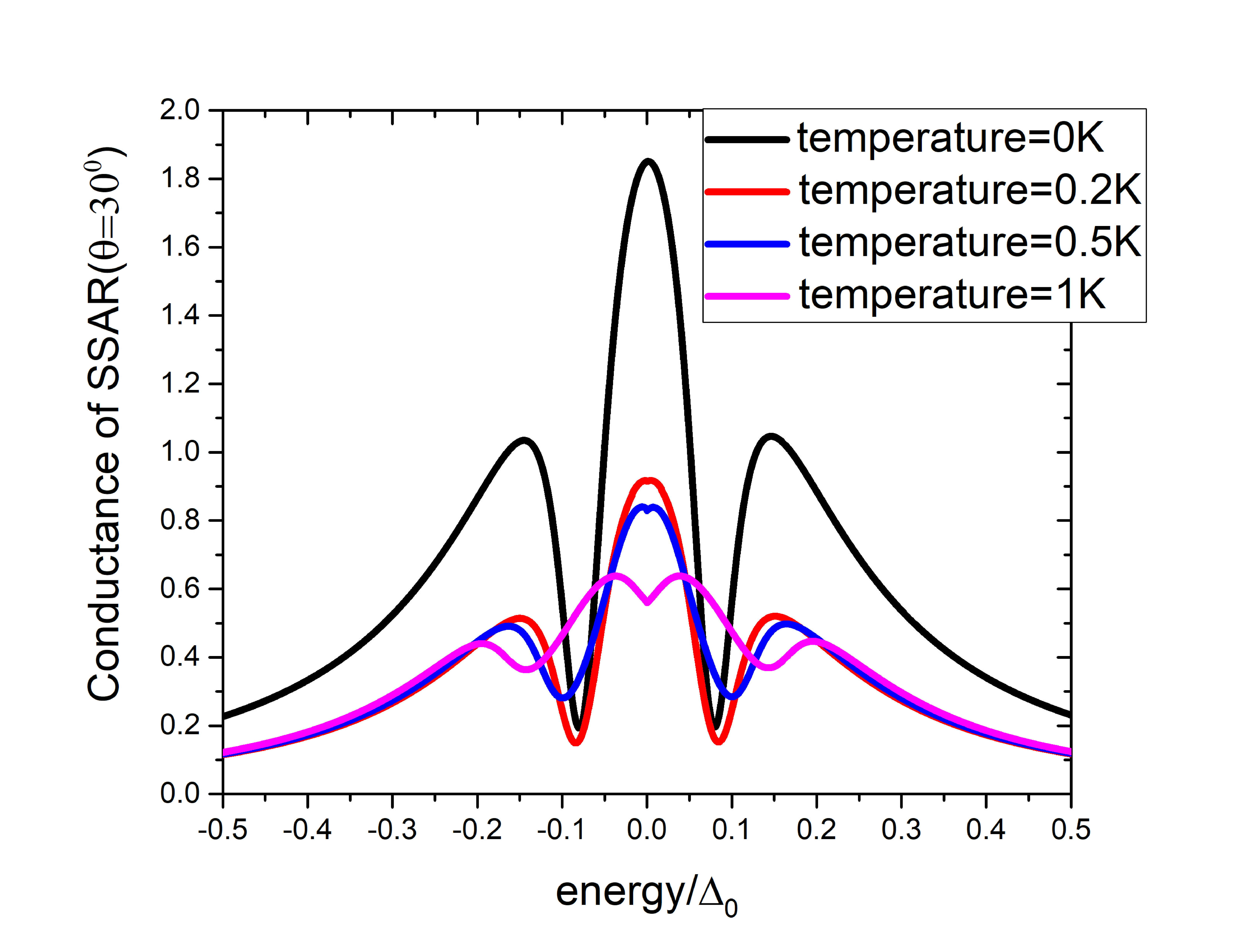}}
 \caption{\raggedright The temperature dependence of the conductance in the SSAR effect when $\theta=30^0$.
 With increasing temperature, the conductance at zero energy is greatly decreased and the width of the conductance line increases.
 In addition, when the temperature is low (1 K), the influence of the first excited states is still observable.
  \label{fig6}}
\end{figure}

The previous analysis in this article was performed at zero temperature.
Here, we study the transport properties under finite temperature conditions in a specific case with $\theta=30^0$.
As shown in Figure ~\ref{fig6}, the conductance decreases greatly with increasing temperature, while the width of the conductance increases with increasing temperature
Due to the thermal broadening under finite temperature, the width of the conductance increases while the maximum value of the conductance decreases.
\\

\section{Shot noise power and the Fano factor }\label{shotnoise}
\begin{figure}
\centering
\scalebox{0.08}[0.08]{\includegraphics[276,88][2832,2484]{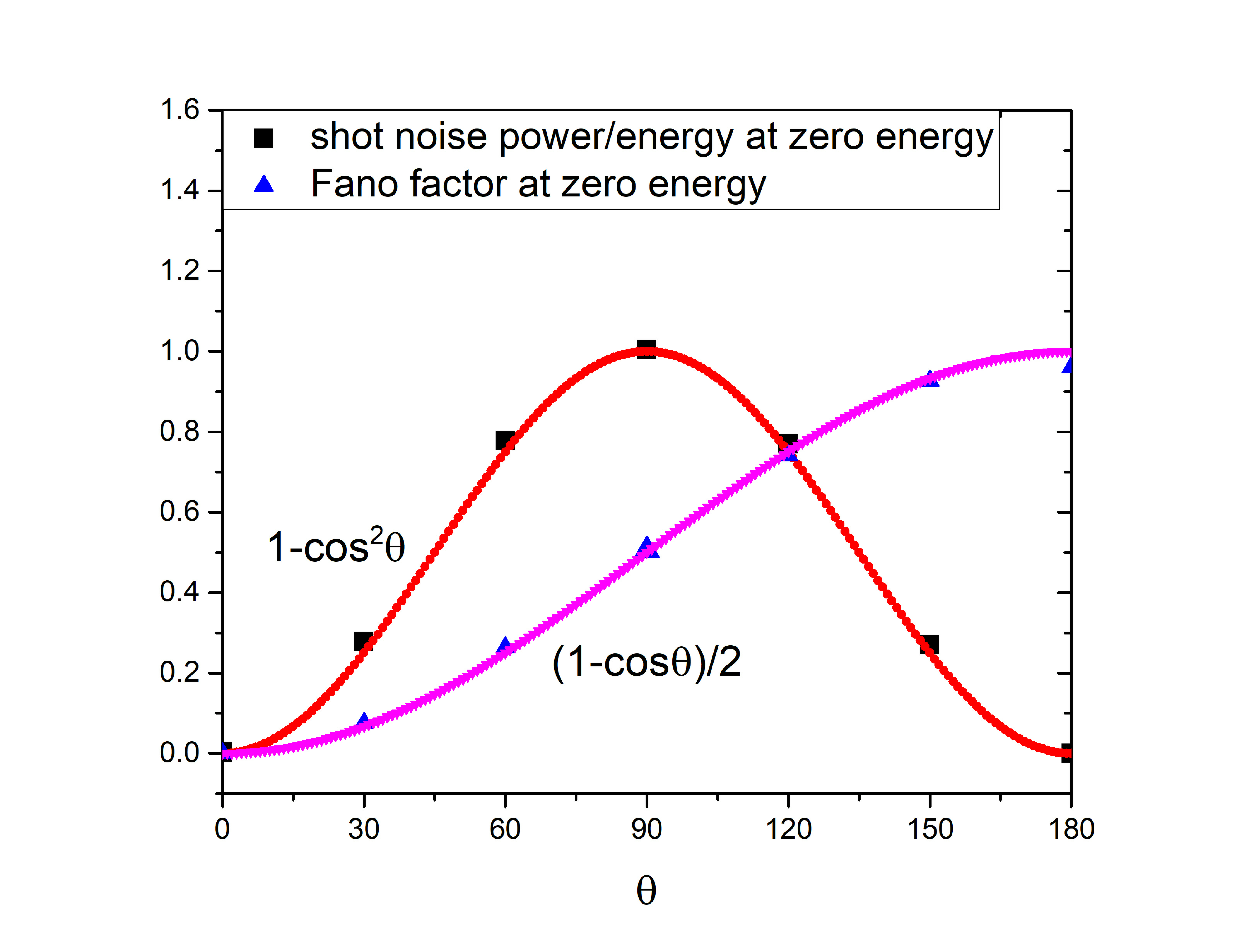}}
 \caption{\raggedright The angular dependence of the shot noise power and the Fano factor at zero energy.
 The shot noise power at zero energy is $1-\cos^2\theta$.
 As opposed to normal Andreev reflection, the Fano factor at zero energy is $(1-\cos\theta)/2$.
  \label{fig7}}
\end{figure}

At zero energy, as shown in Figure ~\ref{fig7}, we find the shot noise is zero due to the full reflection of the electrons when $\theta=0^0$ .
In the noncollinear case, the shot noise power at zero energy will first increase (not more than $90^0$) and then decrease later (more than $90^0$) with increasing $\theta$.
However, the Fano factor(F) increases monotonically with increasing $\theta$, which is very different from normal Andreev reflection\cite{KLAPWIJK19821657,0034-4885-63-10-202,Tinkham}, where F equals two\cite{BLANTER20001}.

Further examining Fig ~\ref{fig7}, we find that F is proportional to $\frac{1-\cos\theta}{2}$ and that S is proportional to $(1-\cos^2\theta)\times$ the energy.
Combined with the number of spin-up electrons that can be reflected as holes is $\frac{1+\cos\theta}{2}$ of the total electron.
We find that the shot noise power is proportional to the number of spin up electrons times the number of spin-down electrons.
\begin{figure}
\scalebox{0.08}[0.08]{\includegraphics[276,88][2832,2484]{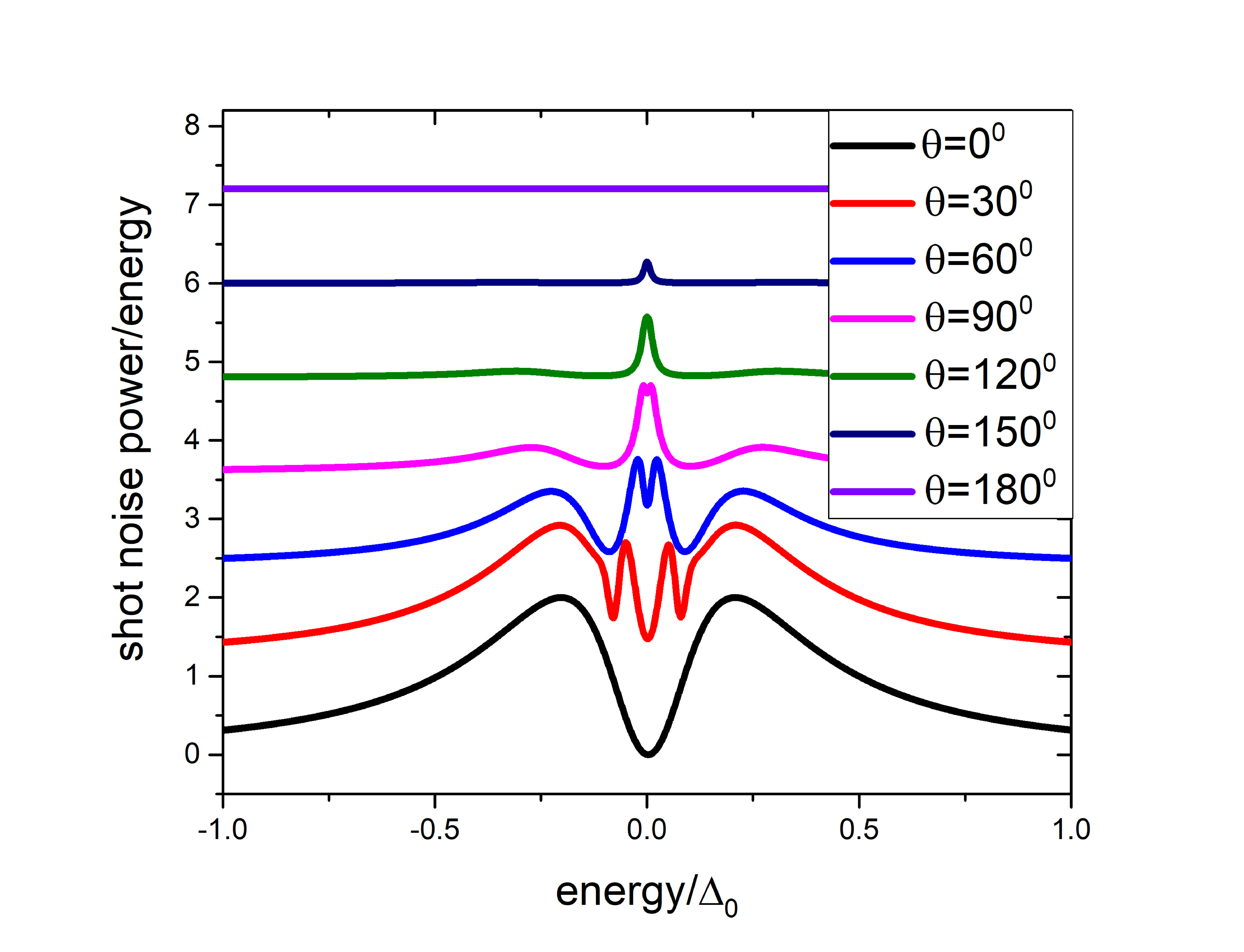}}
 \caption{\raggedright The angular dependence of the shot noise power in the SSAR effect.
 When $\theta=30^0$, $60^0$, $90^0$, $120^0$, $150^0$, and $180^0$, there are shifts of 1.2, 2.4, 3.6, 4.8, 6, and 7.2, respectively.
 When $\theta=0^0$, the shot noise power line has one valley with a minimum value of 0 at zero energy.
However, when $\theta \ne 0$, the shot noise power has more peaks.
In addition, the width of the valley is smaller with increasing $\theta$.
  \label{fig8}}
\end{figure}

Let us focus on the formula of the shot noise power, Eq. (\ref{shotnoise-ev>0}).
As mentioned before, all of the spin-up electrons can be reflected as holes with the same spin while the spin-down electron can be reflected as themselves at zero temperature.
The term that contains $(r_{eh(he)}^+a^+_{h(e)}(E)r_{hh(ee)}a_{h(e)}(E))$ should be zero at zero temperature.
Therefore, the shot noise power can be simplified as
$Sp=\frac{2e^3V}{h}[(r_{ee}^+a^+_e(E)r_{ee}a_e(E))(r_{eh}^+a^+_h(E)r_{eh}a_h(E))+(r_{he}^+a^+_e(E)r_{he}a_e(E))(r_{hh}^+a^+_h(E)r_{hh}a_h(E))]$.
In other words, the shot noise power is proportional to the probability of an electron being reflected as a hole times the probability of an electron being reflected as itself at zero energy.
The Fano factor is equal to $\frac{(1-\cos\theta)}{2}$.

Now, let us look at the shot noise power and the Fano factor at non-zero energy.
As shown in Fig ~\ref{fig8}, we find that, in the collinear case with increasing absolute energy, the shot noise power is first increases and then decreases.
However, the Fano factor always increases with increasing absolute energy.

\begin{figure}
\scalebox{0.08}[0.08]{\includegraphics[276,88][2832,2484]{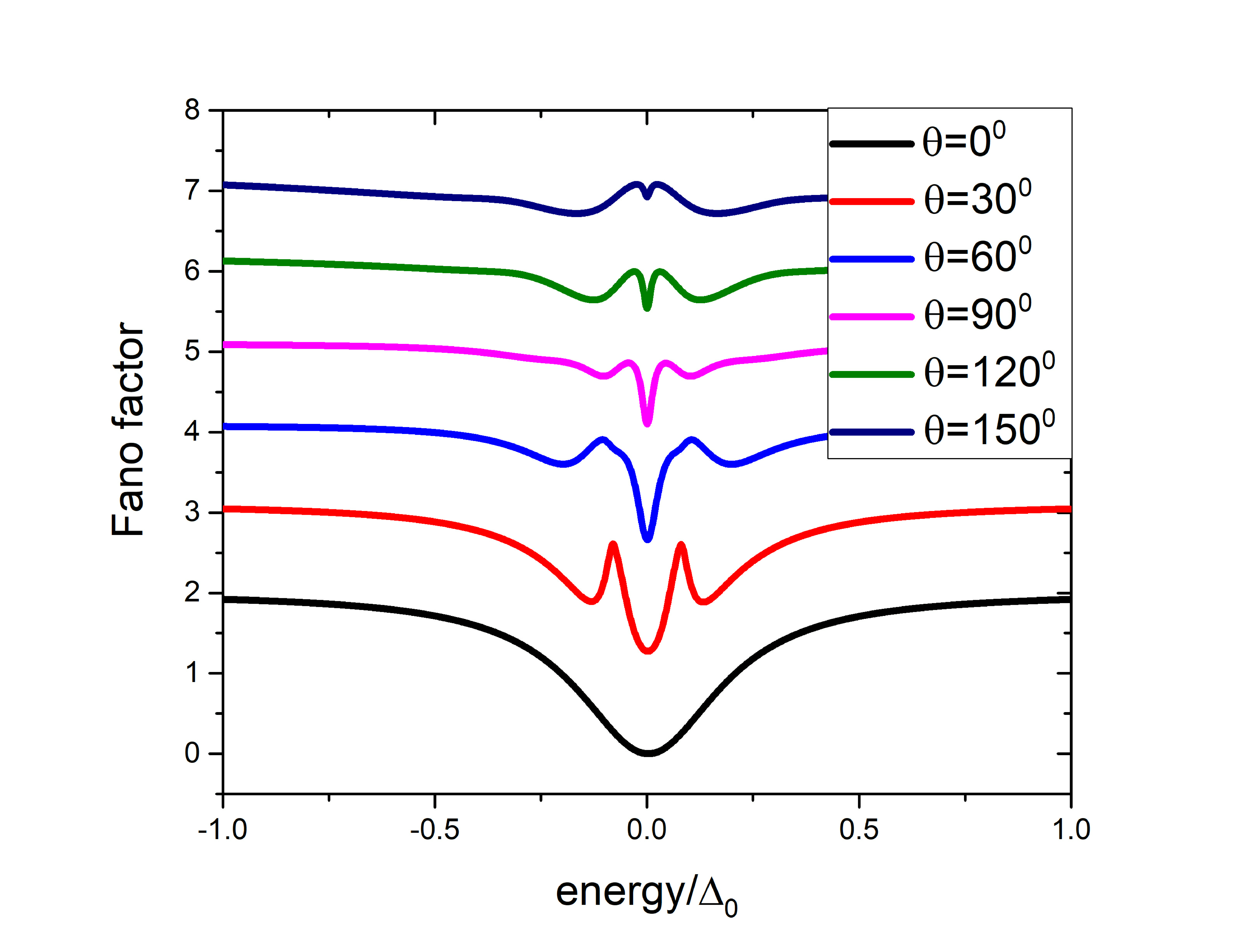}}
 \caption{\raggedright The angular dependence of the Fano factor in the SSAR effect.
 When $\theta=30^0$, $60^0$, $90^0$, $120^0$, and $150^0$, there are shifts of 1.2, 2.4, 3.6, 4.8, and 6, respectively.
 The Fano factor is proposed to be two minus the conductance..
  \label{fig9}}
\end{figure}

In the noncollinear case, with increasing absolute energy, the shot noise power can first increase and then decrease to form multiple peaks, i.e., for$0^0<\theta\leqq90^0$.
However, the shot noise power always decreases with increasing absolute energy when $90^0\leqq\theta<180^0$.
In addition, as shown in Fig ~\ref{fig9}, the Fano factor can first increase and then decrease to form three valleys.

This can be explained by the probability of an electron being reflected as a hole.
The shot noise power is always proportional to the probability that an electron is reflected as a hole times the probability that an electron is reflected as itself.
In the collinear case, the probability that an electron is reflected as a hole is always reduced from 1 to 0 with increasing absolute energy.
Therefore, the shot noise power is increased when the probability isreduced from 1 to 0.5 and is decreased when the probability is reduced from 0.5 to 0.
In the noncollinear case, similar to the conductance, this probability will first induce, then increase and finally reduce.
Note that the shot noise power increases monotonically as the probability increases when the probability is large than 0.5, while the shot noise power decreases monotonically as the probability increases when the probability id less than 0.5.
Therefore, the shot noise power will form a greater number of valleys and peaks in noncollinear case.

Moreover, the Fano factor is proportional to the probability that an electron is reflected as itself.
In collinear case, this probability always increases from 1 to 0 with increasing absolute energy.
Hence, the Fano factor is always increases monotonically as the absolute energy increases in this case.
However, similar to the conductance, the Fano factor will first increase then decrease and increase later with increasing absolute energy in the noncollinear case.

In addition, as mentioned before, the first excited states will influence the probability that an electron is reflected as a hole.
This means the first excited states also have an influence on the shot noise power and the Fano factor.
In more detail, the first excited states can increase the Fano factor and reduce the shot noise power at zero energy when $0^0<\theta<90^0$.
However, when $90^0<\theta<180^0$, the first excited states can reduce the Fano factor and increase the shot noise power at zero energy.
Moreover, the first excited states can make the two original peaks be closer due to the narrowing width of the conductance.
Further, the first excited states can cause more peaks due to the new peak of the conductance induced by the first excited states.
\\

\section{Summary}
We built a model to study the charge transport property of the SSAR effect.
Considering both the MZM and the excited states, we studied the conductance and the shot noise of the noncollinear SSAR effect using Green's function combined with scattering theory.
First, we numerically calculated the angular dependence of the conductance in the SSAR effect.
In our result, the influence of the first excited states of the vortex core on the conductance cannot be ignored when $\theta \ne 0^0$ or $180^0$.
Second, we found that the first excited states can increase the conductance when $\theta<90^0$ and reduce the conductance when $\theta>90^0$ at zero energy.
Next, we focused on the energy of the first excited states and the MZM to separateisolate the contribution of the first excited states.
We found that decreasing the coupling between the vortex core and the STM tip can reduce the influence of the first excited states.
Third, we calculated the influence of the temperature under finite temperature conditions.
With increasing temperature, the maximum of conductance decreases;
however, the width of conductance increases.
At low temperatures, the influence of the first excited states is also obvious.
Finally, we studied the shot noise power and the Fano factor of the SSAR effect.
We found that, at zero energy, the shot noise power is $(1-\cos^2\theta)\times$ the energy.
At this time, the Fano factor is $\frac{1-\cos\theta}{2}$.
This is very different from normal Andreev reflection where the Fano factor is two\cite{BLANTER20001}.
In addition, with increasing $\theta$, the shot noise power is also influenced by the first excited states of the vortex state.
These transport properties can provide more information concerning the detection of MZM via the SSAR effect.
\\
\section{Acknowledgments}
This work is supported by the National Natural Science
Foundation of China (No.11674028, No.61774017, No.11734004 and No.21421003) and National Key Research and Development Program of China(Grant No. 2017YFA0303300).
\bibliography{SSAR}

\end{document}